\documentclass[11pt]{article}
\def\fnote#1#2{\begingroup\def\thefootnote{#1}\footnote{#2}\addtocounter
{footnote}{-1}\endgroup}
\usepackage{graphicx}
\usepackage{epstopdf}

\begin{document}

\hfill{UTTG-10-10}

\vspace{36pt}

\begin{center}
{\large {\bf {Pions in Large-$N$ Quantum Chromodynamics}}}

\vspace{36pt}
Steven Weinberg\fnote{*}{Electronic address:
weinberg@physics.utexas.edu}\\
{\em Theory Group, Department of Physics, University of
Texas\\
Austin, TX, 78712}

\vspace{30pt}

\noindent
{\bf Abstract}
\end{center}

\noindent
An effective field theory of quarks, gluons, and pions,  with the number $N$ of colors treated as large, is proposed as a basis for calculations of hadronic phenomena at moderate energies.  The qualitative consequences of the large $N$ limit are similar though not identical to those in pure quantum chromodynamics, but because constituent quark masses appear in the effective Lagrangian, the `t Hooft coupling in the effective theory need not be strong at moderate energies.  To leading order in $1/N$ the effective theory is renormalizable, with only a finite number of terms in the Lagrangian.

\vfill

\pagebreak

The success of quantum chromodynamics (QCD)  in accounting for processes like electron--positron annihilation into hadrons at high energy shows that it is the correct theory of strong interactions, but it has been difficult to use QCD to account for the wide variety of hadronic phenomena at moderate energies.  

On one hand, the suggestion[1]  to consider QCD in the limit of a large number $N$ of colors, with the gauge coupling $g$ vanishing in this limit as $1/\sqrt{N}$,  has had remarkable success in reproducing {\em qualitative} features of strong interaction phenomena.  But it has not led to much quantitatively, presumably because the `t Hooft coupling $\tilde{g}\equiv g\sqrt{N}$ is not small at moderate energies.  Indeed, with the $u$ and $d$ quark masses negligible, the $N$-independent masses of mesons like the $\rho$ can only be of the order of the integration constant $\Lambda_{\rm QCD}$ in the renormalization group equation for $\tilde{g}$, so inevitably $\tilde{g}$ cannot be small at these meson masses.

Alternatively, it is possible to introduce constituent quark masses into QCD by taking chiral symmetry breaking in account in an effective field theory of quarks, gluons, and pions, so that hadrons (other than the pion) can get their mass mostly from the constituent quark masses.  In consequence,  gluon couplings in the effective theory need not be strong at moderate energies.  This fits in well with the observed pattern of hadron masses, such as the fact that the average of the masses of the nucleon and $\Delta(1238)$ is not very different from $3/2$ the $\rho$ and $\omega$ masses.    Such an effective field theory was briefly mentioned in [2], and proposed and developed in some detail (for the three-flavor case) by 
 Manohar and Georgi[3],  But here there is a different problem: In effective field theories we generally must include every one of the infinite number of interactions satisfying relevant symmetries, all of them presumably  important at moderate energies, so that the theory can only be used at low energies.

I  suggest that by combining these two approaches the difficulties of each can be avoided.  To leading order in $1/N$, the effective field theory of quarks, gluons, and pions is effectively renormalizable, with only a finite number of terms in the Lagrangian needed to absorb all infinities.  Such an effective field theory, with a small value of the `t Hooft coupling at moderate energies, may explain why the naive quark model works so well.
Since the pion is already in the Lagrangian, it is not even necessary for the QCD coupling to be strong at relatively low energies, though it must still be strong at very large distances to keep color trapped.

The effective Lagrangian  is taken as\footnote{Unfortunately, the anomaly that breaks the chiral $U(1)$ symmetry of the Lagrangian with $m_u=m_d=0$ disappears in the large $N$ limit[4], leading to the presence of a light pseudoscalar particle  $\varphi$, a mixture of the $\eta$ and an $SU(3)$ singlet $\eta'$.  The $\varphi$  mass is shown in [4] to vanish as $1/N$ for $N\rightarrow \infty$, so for the consistency of the large $N$ approximation, the field  of this light pseudoscalar presumably should in principle be included in the effective field theory.  It is a failing of the large $N$ approximation that in the real world, with $N=3$, the $\eta$ and $\eta'$ are not particularly light.  Possibly there is some reason why the $\varphi$ mass, though of order $1/N$, is not small.  Here we will not include the $\varphi$ in the effective field theory, but  all of the combinatoric arguments here regarding pions would apply also to $\varphi$ mesons, if they were included in the effective Lagrangian.}
\begin{eqnarray}
{\cal L}_{\rm eff}&=&-\frac{1}{4g^2}{\rm Tr}\left\{F_{\mu\nu}F^{\mu\nu}\right\}-\frac{1}{g^2}\Big(\overline{\psi}(D^\mu\gamma_\mu+m)\psi\Big)\nonumber \\ &&
-\frac{F_\pi^2}{2}{\cal D}_\mu\vec{\pi}\cdot{\cal D}^\mu\vec{\pi}-\frac{2ig_A}{g^2}\Big(\overline{\psi}\gamma_5\gamma^\mu\vec{t}\psi\Big)\cdot {\cal D}_\mu\vec{\pi}\nonumber\\
&&-c_1\left({\cal D}_\mu\vec{\pi}\cdot {\cal D}^\mu\vec{\pi}\right)^2-c_2\left({\cal D}_\mu\vec{\pi}\cdot {\cal D}_\nu\vec{\pi}\right)\,\left({\cal D}^\mu\vec{\pi}\cdot {\cal D}^\nu\vec{\pi}\right)
\;.
\end{eqnarray}
Here $\psi$ and  $\vec{\pi}$ are the quark isodoublet and the pion isovector, re-scaled by multiplying the canonically normalized fields by factors $g$  and $1/F_\pi$, respectively.  Both $g$ and $1/F_\pi$ are taken to go as $1/\sqrt{N}$ for  large $N$ if we hold $\Lambda_{\rm QCD}$ fixed.    Also, $F_{\mu\nu}$ is the re-scaled gluon field strength tensor,  an  $N\times N$ matrix:
$$F_{\mu\nu}\equiv \partial_\mu A_\nu-\partial_\nu A_\mu-i[A_\mu,A_\nu]\;,$$
where $A_\mu$ is $g$ times the canonically normalized  gluon vector potential matrix.  Also, $\vec{t}$ is the quark isospin matrix;  $D_\mu\psi$ is the gauge- and chiral-covariant derivative of the quark field
$$
D_\mu\psi\equiv \left[\partial_\mu-iA_\mu+2i\vec{t}\cdot\frac{\vec{\pi}\times\partial_\mu\vec{\pi}}{1+\vec{\pi}^2}\right]\psi\;;
$$
and  ${\cal D}_\mu\vec{\pi}$ is the chiral-covariant derivative of the pion field
$$
{\cal D}_\mu\vec{\pi}\equiv\frac{\partial_\mu\vec{\pi}}{1+\vec{\pi}^2}\;.
$$
The constituent quark mass matrix $m$ and the axial coupling  $g_A$  are  unconstrained by chiral symmetry, and assumed to be $N$-independent.  (Sum rules have been used in the large $N$ limit to show that quarks  have axial coupling $g_A\simeq 1$[5].) The parameters  $c_1$ and $c_2$  are coefficients of order $N$.

With or without the last two terms in (1), this Lagrangian will clearly reproduce the usual soft-pion theorems of chiral symmetry at low energy, and as we will see, it also gives most of the usual qualitative results[1] of the $1/N$ approximation at moderate energies.  The last two terms in (1) will be needed to cancel ultraviolet divergences
to leading order in $N$.  

Let us consider a process involving some mesons and perhaps also glueballs.  As is well known[1], if we ignore the pions  and keep only the terms for quarks and gluons in (1), the leading connected diagrams  will consist of a single quark loop surrounding  a planar mesh of gluon lines, with insertions of operators in the quark and gluon lines representing the emission and absorption of mesons (other than  pions) and glueballs.  (See Figure 1A.) 
\begin{figure}[t]
	\begin{center}
	\includegraphics[scale=1]{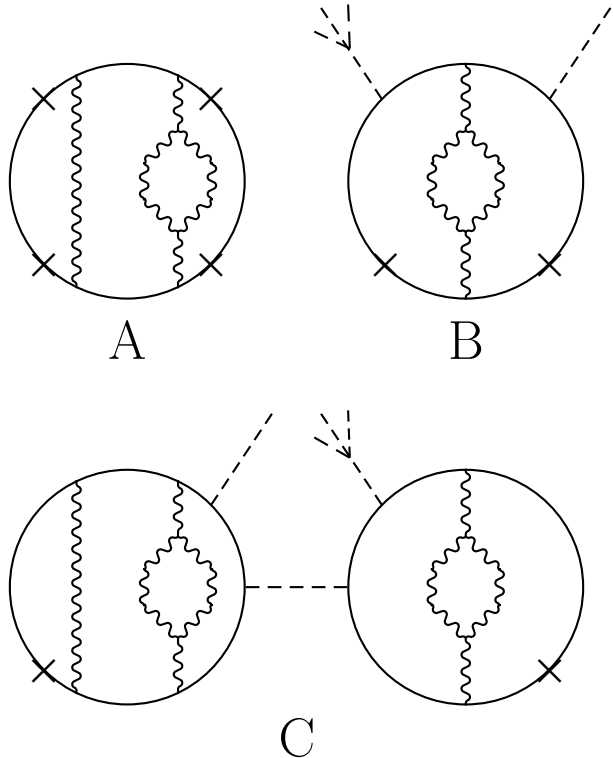}
	\end{center}
	\caption{Some diagrams of leading order in $1/N$ for processes involving pions and other mesons.  Plain lines indicate quarks; wavy lines indicate gluons; dashed lines indicate pions; and crosses indicate insertions of quark bilinears.}
\end{figure}
   Such diagrams make a contribution of order $N$. (This assumes that the operators representing mesons and glueballs are constructed as bilinears  in the {\em un}-rescaled quark and $F_{\mu\nu}$ fields.   For the moment we are ignoring the $N$-dependent factors needed in these insertions to 
give the initial and final states created by these operators the proper normalization, because such factors depend only on the process considered, and hence do not affect the relative contributions of different diagrams for a given process.  These insertions also must include form-factors, taken from the solution of the quark-antiquark bound state problem.)     

Now, suppose we  include  virtual pions, increasing the total number of internal pion and quark lines by $\Delta I$ and increasing the total number  of vertices by $\Delta V$.    Since pions have no color, this does not change the number of index loops, so the  change $\Delta\chi$ in the number $\chi$ of factors of $N$ contributed to a connected graph will be $\Delta\chi=\Delta V-\Delta I$.
But the total number of loops in a connected diagram is $L=I-V+1$, so
$\Delta L=\Delta I- \Delta V$, and thus $\Delta\chi=-\Delta L$.
The dominant connected diagrams will thus be those with $\Delta L=0$.    In other words, these are diagrams  with a single quark loop surrounding a planar mesh of gluon lines, in which the pions add no additional loops, and therefore can only form trees attached to the quark loop. (See Figure 1B.)  Here $\Delta \chi=0$, and such diagrams therefore make contributions of order $N$, just as without pions.

There is, however, a complication here:  The addition of pion lines to an unconnected diagram with $C>1$ separate connected parts may yield a connected diagram --- that is, one with a single connected component.  In this case,  we have 
$\Delta L=\Delta I-\Delta V+\Delta C$, where $\Delta C\leq 0$ is the change in the number of connected components produced by adding the pion lines, which if the graph with pions added is to have a single connected component must be $\Delta C=1-C$.  Thus  the change in the number of factors of $N$ is $
\Delta\chi=\Delta V-\Delta I=-\Delta L+1-C$.  
The leading diagrams for a given $C$ will again  be those with $\Delta L=0$, and now will have $\Delta \chi=1-C$.  
But if the leading graphs without pion lines have $C$ connected components, they are of order $N^C$; that is, they have $\chi=C$.  Hence the leading connected graphs with pions added will have $\chi+\Delta \chi=1$, and so will again be of order $N$.
In contrast to the usual version of large $N$ QCD, in the effective field theory  the leading connected graphs can have any number of quark loops, each surrounding a planar mesh of gluon lines, but connected  in a tree by single pion lines, not gluon lines. (See Figure 1C.)  This allows some ``Zweig-rule-forbidden'' processes in leading order. One can have transitions between $\bar{u}u$ and $\bar{d}d$ mesons by having one meson destroyed at one quark loop and the other created at another quark loop, but only if these mesons have the quantum numbers of the pion.  In leading order there are no pion lines connecting quark lines within a single quark loop, so pion exchange has no effect on the spectrum of mesons, other than those with the quantum numbers of pions.  Also, to leading order the renormalization group equation for the `t Hooft coupling in the effective theory is the same as in QCD, but the integration constant $\Lambda$ in the solution of this equation may be smaller, giving a smaller coupling at any given energy.

The same analysis applies to the case $C=0$.  That is, a tree graph consisting solely of pion lines with vertices given by the purely pionic  terms in Eq.~(1) makes a contribution to purely pionic processes of order $N$, just like diagrams for the same processes with one or more quark loops.

Unlike the usual experience with  effective field theory[2,3], for large $N$ we have no additional ultraviolet divergences due to loops including pion fields.  The attachment of vertices proportional to ${\cal D}_\mu\vec{\pi}$
 to a quark loop does introduce new ultraviolet divergences, but in such diagrams the pion field acts just as a classical external field, so the ultraviolet divergences are limited.  There are  logarithmic divergences from graphs with four new vertices in the quark loop, whose form is constrained by chiral symmetry so that they can be canceled\footnote{If we did include the field $\varphi$ of an isoscalar pseudoscalar pseudo-Goldstone boson in the theory, we would also have to include counterterms
$$ -c_3\left(\partial_\mu\varphi\partial^\mu\varphi\right)^2-c_4\left(\partial_\mu\varphi\partial^\mu\varphi\right)\left({\cal D}_\mu\vec{\pi}\cdot {\cal D}^\mu\vec{\pi}\right)-c_5\left(\partial_\mu\varphi\partial_\nu\varphi\right)\left({\cal D}^\mu\vec{\pi}\cdot {\cal D}^\nu\vec{\pi}\right)\;,$$
in addition to  kinematic and axial vector coupling terms for $\varphi$.} by the terms in (1) proportional to $c_1$ and $c_2$.  There are  quadratic divergences from graphs with two new vertices in the quark loop, that can be canceled by renormalization of $F_\pi$.  These graphs also produce logarithmic divergences, that as remarked in [2] can be canceled by redefinition of the pion field.  Finally, gluon corrections to a vertex inserted in a quark line produce logarithmic divergences that can be canceled by renormalization of $g_A$.  Thus {\em in the large $N$ limit the Lagrangian (1) describes what is in effect a renormalizable theory.}  Terms in the Lagrangian with more quark or $F_{\mu\nu}$ field factors and/or more derivatives are not  needed for renormalization in leading order in $N$, so such terms may be  taken to have coefficients with sufficient powers of $1/N$ so that they do not contribute in leading order.

The dominant graphs remain of order $N$ (or $N^2$, for reactions involving only glueballs) if the insertions in quark and gluon lines that we make to represent the emission and absorption of mesons (other than pions) and glueballs are bilinear in un-rescaled fields, and hence of order $N$ when expressed in terms of re-scaled quark and gluon fields, like the terms in (1).  In particular, propagators of these operators are of order $N$ for mesons (as also for the re-scaled pion field) and of order $N^2$ for glueballs.  But an operator that is properly normalized to produce physical states must have a propagator whose residues at one-particle poles are $N$-independent, so to form  properly normalized operators for creating and destroying pions and other mesons we must include an additional factor proportional to $1/\sqrt{N}$, while the properly normalized operators for glueballs must include an additional factor $1/N$.    
The amplitude for reactions whose initial and final states contain respectively  $M\geq 1$ and $M'\geq 1$ pions or other mesons and $G\geq 0$ and $G'\geq 0$ glueballs is then of order $N^{1-M/2-M'/2-G-G'}$, as in the usual case without pions[1].  Since pions count here the same as other mesons, the usual arguments show that the singularities of scattering amplitudes for mesons and glueballs consist solely of meson poles in various channels.

Finally,  let us consider the leading connected terms in the interaction of the $N$ quarks making up a baryon.  Witten[6] has shown that in pure QCD, without pions, the leading contributions to a connected graph involving $n$ quarks is of order $N^{1-n}$, but the number of ways of selecting these $n$ quarks from the $N$ quarks in the baryon is $N!/(N-n)!n!$, which for $N\gg n$ goes as $N^n$, so that the sum of these connected graphs is of order $N$.  In the effective theory including pions, we have to take into account the possibility of forming a connected graph from a disconnected diagram with $C$ separate connected parts, linking them together with pion lines.   
\begin{figure}[t]
	\begin{center}
	\includegraphics[scale=1]{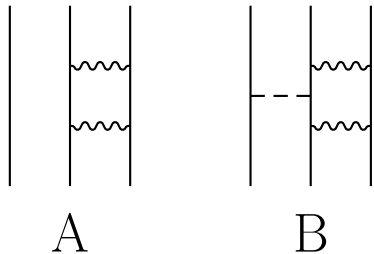}
	\end{center}
	\caption{Some diagrams of leading order in $1/N$ for baryon structure.  Notation same as Fig. 1.}
\end{figure}
 First consider the $N$-dependence of such a disconnected diagram before we add the pion lines.  (See Figure 2A.) If the $r$th connected part involves $n_r$ quark lines,  then the total contribution of such parts is of order 
$$ \prod_{r=1}^C N^{1-n_r}\; \times \; \frac{N!}{C!(N-\sum_r n_r)!\prod_r n_r!}\rightarrow \frac{N^C}{C!\prod_r n_r!}\;,$$
where the first factor is the contribution of the $C$ connected parts, and the second factor is the number of ways of selecting the quarks in these $C$ connected parts.    Just as we saw in the meson case, the addition of pion lines to give a connected diagram supplies an additional factor $N^{1-C-\Delta L}$, where $\Delta L$ is the increase in the number of loops, so the leading graphs are those in which the addition of pions does not increase the number of loops, and these graphs are of order $N$, as in pure QCD.  (See Figure 2B.)

This picture raises issues of double counting of baryons.\footnote{Ref. [3] has already shown a way that binding a second pion can be avoided.}    The purely pionic part of the Lagrangian (1) may have skyrmion solutions, with masses of order $N$[6], in addition to the $N$-quark states described above.   Indeed, the last two terms in (1) are just the sort needed to stabilize the skyrmion.  But although in this theory purely pionic interactions are correctly described at low energy in the tree approximation by the purely pionic terms in (1), this is not true at the moderate energies of order $\Lambda_{\rm QCD}$ probed in the structure of skyrmions.  At such energies, to leading order in $N$ we must also take into account quark loops, each surrounding a planar mesh of gluon lines, which can take the place of vertices in a tree of pion lines.  Nothing is known about the existence of skyrmion solutions when such quark loops are taken into account.

This work leaves open several questions:  Can (1) be derived from QCD by some process of ``integrating out'' degrees of freedom?  If so, what is the relation between the integration constants $\Lambda$ for the `t Hooft couplings in QCD and the effective theory?  And  will this effective field theory provide a basis for practical calculations of hadronic phenomena at moderate energies?

I am grateful for valuable conversations with J. Distler, W. Fischler,  V. Kaplunovsky, and J. Meyers.  This material is based upon work supported by the National Science Foundation under Grant Numbers PHY-0969020 and PHY-0455649 and with support from The Robert A. Welch Foundation, Grant No. F-0014.

\vspace{10pt}

\begin{center}
{\bf ----------}
\end{center}

\vspace{10pt}

\begin{enumerate}
\item G. 't Hooft, Nucl. Phys. {\bf B75}, 461 (1974).
\item S. Weinberg, Physica {\bf 96A}, 327 (1979).
\item A. Manohar and H. Georgi, Nucl. Phys. {\bf B234}, 189 (1984).
\item E. Witten, Nucl. Phys. B {\bf 149}, 285 (1979); {\bf 156}, 269 (1979); G. Veneziano, Nucl. Phys. B {\bf 159}, 213 (1979).
\item S. Weinberg, Phys. Rev. Lett. {\bf 65}, 1181 (1990).
\item E. Witten, Nucl. Phys. {\bf B160}, 57 (1979).
\end{enumerate}

\end{document}